\documentclass[showpacs,preprintnumbers,amsmath,amssymb,12pt,floatfix,epsfig]{revtex4}
\usepackage{dcolumn}
\usepackage{bm}
\usepackage[dvips]{color}
\usepackage{graphicx}

\baselineskip=24ptplus.5ptminus.2pt \vspace*{0.1 in} \large

\def\prl#1{Phys.\ Rev.\ Lett.\ {\bf #1}}
\def\pr#1{Phys.\ Rev.\ {\bf #1}}
\def\np#1{Nucl.\ Phys.\ {\bf #1}}
\def\pl#1{Phys.\ Lett.\ {\bf #1}}
\def\prt#1{Phys.\ Reports.\ {\bf #1}}
\def\jpsj#1{J.\ Phys.\ Soc.\ Jpn.\ {\bf #1}}
\def\be{\begin{equation}}
\def\ee{\end{equation}}
\def\Be{\begin{eqnarray}}
\def\Ee{\end{eqnarray}}
\def\ba{\begin{array}}
\def\ea{\end{array}}
\begin{document}
\title{\bf The pseudo scalar form factor of the nucleon, the sigma-like term, and
the $L_0^+$ amplitude for charged pion electro-production near
threshold}
\author{Myung Ki Cheoun$^{1)}$ \footnote{Corresponding Author, e.mail : cheoun@ssu.ac.kr} and K. S. Kim$^{2)}$}
\affiliation{\it 1) Department of Physics, Soongsil University,
Seoul, 156-743, Korea}
\affiliation{\it 2) School of Liberal Arts
and Science, Hankuk Aviation University, Koyang, 412-791, Korea}

\begin{abstract}
The pseudo scalar form factor, which represents the pseudo scalar
quark density distribution due to finite quark masses on the
nucleon, is shown to manifest itself with the induced pseudo
scalar form factor in the $L_0^+$ amplitude for the charged pion
electro-production. Both form factors show their own peculiar
momentum dependence. Under the approximation on which the
Goldberg-Treimann relation holds, a sum of both form factors'
contributions accounts for the t-channel contribution in the
charged pion electro-production near threshold.
\end{abstract}

\maketitle

---------------------------------------------------------

An explicit chiral symmetry breaking (ECSB) effects originated
from quark masses can be realized as the scalar form factor of the
nucleon $\sigma (t)$ in the following way
\begin{eqnarray}
{\langle  N ( p_2 ) \vert {\hat m} {( {\bar u} (0) u (0) + {\bar
d} (0) d (0) )} \vert N (p_1) \rangle} = \sigma (t) {{\bar u}(
p_2) u (p_1)}~,\\ \nonumber ~{\hat m} = (m_u + m_d) / 2~,~t =
{(p_2 - p_1)}^2~. \label{sca}
\end{eqnarray}
A few experimental results for the scalar form factor are only
available at $t = 2 m_{\pi}^2$, so called Cheng-Dashen point. They
were extracted from the $\pi - N$ scattering through a low energy
theorem \cite{Ch71}. Old analysis yield 50$\sim$70 MeV, while more
recent analysis show 71$\sim$90 MeV. Detailed references for
experimental results can be found at ref. \cite{Sch04}.

The so called $\sigma_{\pi N}$ term is defined as a value of
$\sigma (t)$ at $t=0$ {\it i.e.} $\sigma_{\pi N} = \sigma (t)
\vert_{t = 0}$, so that it has been investigated mostly by
theoretical ways because of its problematic extrapolation to
$t=0$. For example, chiral perturbation theory and dispersive
analysis \cite{Ga91,Fuc03-2} showed $\Delta \sigma = \sigma(t)
\vert_{ t = 2 m_{\pi}^2 } - \sigma (t) \vert_{t=0}$ = 15$\sim$17
MeV. Modern values of $\sigma_{\pi N}$, therefore, turned out to
be about 10 MeV larger than old ones, and such a large value of
$\sigma_{\pi N}$ gives rise to many interesting problems in the
relevant fields, for instance, the strangeness content of the
nucleon, searches of the Higgs boson and so on \cite{Sch04}.

Likewise one can also define the pseudo scalar (PS) form factor
$\Pi (t)$ standing for the PS quark density on the nucleon
\begin{equation}
{\hat m} {\langle  N ( p_2 ) \vert  { {\bar q} (0) \gamma_5 \tau_a
q (0) } \vert N (p_1) \rangle} = \Pi (t)  {{\bar u} (p_2)}
\gamma_5 \tau_a u (p_1)~.
\end{equation}
With the help of the interpolating pion field $\Phi_a (x)$ defined
by $ {\langle  0 \vert \Phi_a (x)  \vert \pi_b (q) \rangle} =
\delta_{ab} e^{-i q x} $ and the definition of coupling constant $
{\langle  0 \vert P_a (0)  \vert \pi_b (q) \rangle} = \delta_{ab}
G_{\pi}$ with $P_a = i {\bar q} \gamma_5 \tau_a q$ and ${\hat m}
G_{\pi} = f_{\pi} m_{\pi}^2$, the PS quark density can be
parameterized as \cite{Fuc03}
\begin{equation}
{\hat m} {\langle  N ( p_2 ) \vert  {{\bar q} (0) \gamma_5 \tau_a
q (0) } \vert N (p_1) \rangle} = {  { m_{\pi}^2 f_{\pi} } \over {
m_{\pi}^2 - t }} G_{\pi N} (t) {{\bar u}  (p_2)} \gamma_5 \tau_a u
(p_1)~.
\end{equation}
Contrary to the scalar form factor, therefore, under the one pion
exchange approximation the PS form factor $\Pi (t)$ is simply
expressed as

\begin{equation}
\Pi (t) = {  { m_{\pi}^2 f_{\pi} } \over { m_{\pi}^2 - t }} G_{\pi
N} (t)~.
\end{equation}
Here $G_{\pi N} (t)$, referred as the pion-nucleon form factor,
can be calculated by the chiral perturbation theory with a given
effective lagrangian as follows
\begin{equation}
G_{\pi N} (t) = g_{\pi N} ( 1 + \Delta_{\pi N} { {t - m_{\pi}^2  }
\over {m_{\pi}^2 }} )~~ with~~ g_{\pi N} = G_{\pi N} ( m_{\pi}^2).
\end{equation}
The $\Delta_{\pi N}$ stems from the violation of the
Goldberger-Treimann(GT) relation {\it i.e.} $\Delta_{\pi N} \equiv
1 - { {g_A M_N} \over {g_{\pi N} f_{\pi} }}~$, and can be
evaluated from a given effective lagrangian up to any order. We
follow the results at ref.\cite{Fuc03,Bern02}, in which an
effective lagrangian up to the 3rd order is considered.

In this work this PS form factor is shown to manifest itself with
the induced PS form factor $G_P(t)$ in the transition amplitude
for the electro-production of charged pions, and consequently may
cause some additional ambiguities in the extraction of the $G_P
(t)$ and the pion form factor $F_{\pi} (k^2)$ from the $L_0^+$
amplitude data. An estimation of the contributions of the $G_P(t)$
and the $\Pi (t)$ to the $L_0^+$ amplitude near pion threshold is
finally suggested with some remarks on previous analytical and
phenomenological formulae for the extraction of the amplitude.

To describe the pion electro-production $\gamma^* (k) +
N(p_{1}){\rightarrow}{\pi}_{a}(q) + N(p_{2})$, where $a$ is a
cartesian isospin, we start from the invariant amplitude ${\cal
M}_a = e \epsilon_{\nu} {\cal M}_{a}^{\nu} = e \epsilon_\nu
\langle N(p_2), \pi_a (q) \vert J^{\nu} (0) \vert N (p_1) \rangle$
with the polarization vector $\epsilon_{\nu}$ of the lepton part.

We exploit the following Green functions $C_a^{\nu}$, $P_a^{\nu}$
and $T_a^{\mu \nu}$ \cite{Sch03}, the nucleon matrix elements of
the time ordered products of the relevant currents {\it i.e.} the
vector current $J^{\nu}$, the axial current $A_a^{\nu}$, and the
PS quark density $P_a$. They are derived from the QCD lagrangian
in the presence of the external electro-magnetic (EM) fields
\begin{eqnarray}
C_a^\nu & = & {\int} {d^{4} x}  {e^{iqx}} \delta (x_0) \langle
p_{2} \vert   [ A_{a}^{0} (x), J^{\nu} (0) ] \vert p_{1} \rangle~,
\\ \nonumber P_a^\nu & = &i {\int} {d^{4} x}  {e^{iqx}} \langle p_{2}
\vert  T ( P_a(x) J^{\nu} (0)) \vert p_{1} \rangle~,
\\ \nonumber T^{\mu \nu}_a &= &i {\int} {d^{4} x} {e^{iqx}} \langle
p_{2} \vert  T ( A_{a}^{\mu} (x) J^{\nu} (0) ) \vert p_{1}
\rangle~, \label{gftn}
\end{eqnarray}
which satisfy the chiral Ward identity, $ q_{\mu} T^{\mu \nu}_a =
{\hat m}P_a^\nu + C_a^\nu$, held for each order in the given
effective lagrangian \cite{Fuc03,Sch03}.

Then the transition current matrix element ${\cal M}_{a}^{\nu}$ is
obtained by the LSZ reduction formula with the interpolating pion
field $\Phi_a (x)$ within the Bjorken-Drell convention as follows
\begin{eqnarray}
{\cal M}_{a}^{\nu} & = & ~\int {d^{4}x}~ {e^{iqx}} ( \partial^2
+m_{\pi}^{2}) {\langle}{p_{2}}{\vert}
T({\Phi_{a}}(x)J^{\nu}(0)){\vert}p_{1}{\rangle} \\
\nonumber & = &  ( { { q^{2} - m_{\pi}^2 } \over { m_{\pi}^2
 f_\pi  } } ) {\int} d^{4} x ~e^{iqx} \langle p_{2} \vert
\partial_\mu T ( { {A_{a}^{\mu}}} (x) J^{\nu} (0) ) -
\delta (x_0) [{A_a^0} (x) , J^\nu (0) ] \vert p_{1} \rangle ~.
\label{manu}
\end{eqnarray}

Finally we obtain the following transition amplitude
\begin{eqnarray}
f_\pi {\cal M}_a^\nu =  [ ( C_a^\nu + q_{\mu} T_{a}^{\mu \nu} ) -
{q^2 \over m_{\pi}^2} ( C_a^\nu + q_\mu T_{a}^{\mu \nu } )]
~\vert_{~q^2 \rightarrow m_{\pi}^2}~.
\end{eqnarray}
By the equal time commutator (ETC) of the axial charge and the
vector current, $C_a^{\nu}$ is reduced to $C^{\nu}_{a}  = {\bar
u}(p_{2}) {I_{a} \over 2} [ G_{A}(t){\gamma^{\nu}}{\gamma_{5}}
 +{{G_{P}(t)} \over 2M_N}{{(k - q)}^{\nu}}{\gamma_{5}}] u(p_{1})$
 with $ I_{a} = i {\epsilon_{a3b}}{\tau_{b}}$,
 where the 2nd term is the induced PS part and
contributes exclusively to the charged pion electro-production and
also to the muon capture processes.

The above result can be also derived from the pre-QCD PCAC
hypothesis \cite{Fuc03} $\partial_{\mu} A^{\mu}_a= m_{\pi}^2
f_{\pi} {\Phi_{a}}$. For example, in our previous calculation
\cite{cheoun}, we have started from a given model lagrangian that
satisfies the PCAC hypothesis. This hypothesis corresponds just to
the inclusion of the EM interactions by the minimal coupling
scheme \cite{Habe01}, so that it may not have any direct relation
to the QCD and may lead to a failure incompatible with the chiral
symmetry as discussed in ref.\cite{Sch03}. But the approach
adopted here is based on the chiral Ward identity from the QCD
lagrangian.

Under the soft pion limit $q^2 \rightarrow 0$, the 2nd parenthesis
part does not contribute, so that one easily obtains the classical
LET result \cite{Meis96} from the 1st parenthesis. But, under the
real pion limit $q_\mu \rightarrow m_{\pi} $, the 2nd parenthesis
can be divided as \cite{Domb73, Schae91}
\begin{equation}
{ i q^0 \over m_{\pi}^2 }  \int {d^{4}x}{e^{iqx}}
{\langle}{p_{2}}{\vert} ~ \delta (x_0) [
\partial_\mu A_a^{\mu}(x) ,  J^{\nu}(0) ]
  {\vert}p_{1}{\rangle} + ~ i q_{\mu}
  \int d^{4} x e^{iqx} \langle p_{2} \vert T ( f_\pi
  \partial^\mu
 \Phi_a (x) J^{\nu} (0) ) \vert p_{1} \rangle ~,
 \label{second}
\end{equation}
where we made use of the chiral Ward identity mentioned above.
Here, if we remind that the $\sigma$ term in $\pi N $ scattering
is given as $ \sigma_{\pi N} = i {\int}{d^{4}x} {\delta}(x_{0})
{\langle}{p}{\vert}
 [ A_a^{0},{\partial}_{\mu} {A}_a^{\mu}(x)]
 {\vert}p{\rangle}
$, we can define the 1 st term as the sigma-like term, which is
denoted as ${\Sigma_a^{\nu}}_{ (\gamma^* \pi^a)} $ to distinguish
from the $\sigma_{\pi N}$ term in the $\pi - N$ scattering, by
replacing $A_a^0$ with $J^{\nu}$
\begin{equation}
 {\Sigma_a^{\nu}}_{ (\gamma^*
\pi^a)} = {\int}{d^{4}x}~{e^{iqx}} {\delta}(x_{0})
{\langle}{p_{2}}{\vert}
 [ {\partial}_{\mu} {A}_a^{\mu}(x), J^{\nu}]
 {\vert}p_{1}{\rangle}~.
\label{sigma}
\end{equation}
The $\sigma_{\pi N}$ is defined by contracting the isospin and
taking $t \rightarrow 0$ limit, but in the ${\Sigma_a^{\nu}}_{
(\gamma^* \pi^a)} $ defined here we keep them because it appears
explicitly at the transition amplitude. Hereafter we drop the
subindex $\gamma^{*} N$. Since its integrated part goes to zero in
the chiral limit, this $\Sigma_a^{\nu}$ term is also an ECSB term.

The 2nd term in eq.(\ref{second}) yields with $q_\mu T_{\mu
\nu}^a$ in the 1st parenthesis of eq.(8)
\begin{equation}
i q_\mu {\tilde T}_a^{\mu \nu}
 = i q^\mu {\int}{d^{4}x}{e^{iqx}}{\langle}{p_{2}}{\vert}
 T ({\tilde A}_a^{\mu}(x)J^{\nu}(0)) {\vert}p_{1}
{\rangle}~,
\end{equation}
where ${\tilde A}_{\mu}^a = A_{\mu}^{a}(x) -
{f_{\pi}}{\partial}_{\mu}{\Phi}^{a}(x) $ is the axial current with
the pion axial current subtracted. As a result, one does not
retain the pion pole structure in $ i q^{\mu} {\tilde T}_{\mu
\nu}^a $ any more. Therefore, our transition amplitude is finally
summarized as follows
\begin{eqnarray}
f_{\pi} {\cal M}_{a}^{\nu}
 = C_{a}^{\nu}+
 i { {q_{\mu}}  } {\tilde T}_{a}^{\mu \nu}
 + {  {i q_0} \over { {m_{\pi}^2  } }} {{\Sigma}_{a}^{\nu}}
~. \label{finalamp}
\end{eqnarray}
This relation is identical to the result of ref.\cite{Schae91},
which is derived from the Ward-Takahasi identity. Since the
$\Sigma_a^{\nu}$ term contribution is reduced to a part of the
t-channel in the Born approximation as shown later on, we have no
difficulties to maintain the gauge invariance.

In this report, we take the spatial part of the $\Sigma_a^{\nu }$
term to be zero following the arguments in ref.
\cite{Kama89,Bern91}. They claim that the commutator $ [
\partial_{\mu} A_a^{\mu} (x) , J^{\nu = i} (0) ] $ in $\Sigma_a^{\nu}$ should
be zero if the model used satisfies the PCAC. The time component
$\Sigma_a^{\nu = 0}$ is detailed in this report.

If one uses the conservation of the vector current, the equal-time
commutator ${[ Q_a^5, J^\mu (y) ]}_{x_0 = y_0} = i \epsilon_{a 3
b} A_b^\mu (y)$ with $ Q_a^5 = \int d^3 x
 A_a^0 (x) $, and its model independent derivative form
 $ [ D_a (x) , J^0 (0) ]  = i \epsilon_{a 3 b} D_b (0) $ with
$D_a (x) =$ $\int d^3 x$ $\partial_{\mu} A_a^{\mu} ( x) $, the
time component of eq.(\ref{sigma}) is reduced to the nucleon
expectation value of the axial current divergence, and
consequently expressed in terms of the PS form factor $\Pi (t)$
\begin{equation}
 {i \Sigma_a^0  }   =  -
\epsilon_{a 3 b} \langle p_2 \vert \partial_\mu A_b^\mu (0) \vert
p_1 \rangle   =  -  \Pi (t) {\bar u} (p_2) \gamma_5 { I_a }u
(p_1), \label{sigmaeq16}
\end{equation}
where we used $\partial_\mu A_b^\mu = {\hat m} P_b = {\hat m} i
{\bar q} \gamma_5 \tau_b q$. The isospin structure $I_a = i
\epsilon_{a 3 b} \tau_b$ leads no contribution of this term to
$\pi^0$ production likewise the $C_a^{\nu}$ term. For the charged
pion production (we distinguish the pion charge as a Greek letter)
this can be rewritten
\begin{equation}
i \Sigma_{\alpha}^0 =  {\hat m} \langle p_2 \vert {\bar q}_i
\gamma_5 { {[\tau_{\alpha} , \tau_3]
  } \over 2} q_i \vert p_1 \rangle ~,
   \label{sigmaeq25}
\end{equation}
which stands for $ {\sqrt 2} {\hat m} \langle p_{2} \vert - {\bar
d} \gamma_{5} u \vert p_{1} \rangle$ for $ \pi^+$ and ${\sqrt 2}
{\hat m} \langle p_{2} \vert  {\bar u} \gamma_{5} d \vert p_{1}
\rangle$ for $ \pi^-$ productions, respectively.

If we take an average value for the spins of the initial and final
nucleons
\begin{equation}
i {\bar \Sigma}_{\alpha}^0 = { 2 \over { \pi }} \sqrt{{ { {( W-
M_N)}^2 - k^2}  \over {4 W M_N} }  } ~ \Pi (t)~ <{  { [
\tau_{\alpha}, \tau_3] } \over 2}    > \label{sigmaeq19}
\end{equation}
and a value of $\Pi (t)$ at the pion threshold $ t_{thre.} = (k^2
- m_{\pi}^2 ) / ( 1 + { { m_{\pi}} \over { M_N}})$, we obtain $i
{\bar \Sigma}_{\alpha}^0 \vert_{thre.} $ = 38.1 MeV at the
$\gamma$ point {\it i.e.} the $k^2 = 0$ point. Direct calculations
by QCD or QCD inspired models could give more systematic
comparison of this quantity.

The neutral PS quark density $\langle p_2 \vert {\bar u} \gamma_5
u + {\bar d} \gamma_5 d\vert p_1 \rangle $ can be also realized if
the singlet currents $J_0^\mu$ and $A_0^\mu$ in the $U(1)_V$ and
the $U(1)_A$ gauges are taken into account. For instance, in weak
pion production one can expect such a quantity. But only the EM
production of the charged pion {\it i.e.} only the charged PS
quantity is considered in this work.

Most experiments for pion production near threshold by EM
processes were carried out to extract the axial mass $M_A$ from
the $E_0^+$ amplitude, and the pion form factor $F_{\pi} (k^2)$
through the induced PS form factor $G_P (t)$ from the $L_0^+$
amplitude \cite{Bern02,Sch91}. But the model dependent terms in
$L_0^+$ amplitude are said to be of a size on which both form
factors could not be distinguished. Moreover only a few data for
this reaction near threshold are available until now
\cite{Choi93,Mainz}.

However our transition amplitude at the pion threshold allows not
only the $G_P(t)$ but also the $\Pi (t)$
\begin{equation}
{{\epsilon_\nu C_a^\nu} \over f_{\pi}} \vert_{pole} + {{i
\epsilon_\nu \Sigma_a^{\nu}}\over {m_{\pi} f_{\pi}   }} = {\bar u}
(p_2) { I_a \over 2} \epsilon_0 [ - { {m_{\pi} }\over { f_{\pi}
}}~ ({ G_P(t) \over {  2 M_N}}  + { {2  \Pi (t)} \over {m_{\pi}^2
} }) ] \gamma_5 u (p_1)~, \label{total}
\end{equation}
where we consider only the pole contribution {\it i.e.} omit the
contact term in the ${\epsilon_\nu C_a^\nu}$ part for further
discussion, but it will be taken into account in the calculation
of the amplitudes. Although both form factors have the pion pole
dominance, their dependence on momentum transfer $t$ is different
and independent of each other if we recollect the following
relation among the form factors, $ 2 M_N G_A (t) + { t \over { 2
M_N}} G_p (t) =  {{2 m_{\pi}^2 f_{\pi} \over { m_{\pi}^2 - t }}}
G_{\pi N} (t) = (2 \Pi (t))$. Actually this relation, a momentum
dependent (or generalized) GT relation, holds even in the 3rd
order lagrangian ${\cal L}_{eff.}^{(3)}$. Detailed forms for each
form factor are taken from ref.\cite{Fuc03,Bern02}
\begin{equation}
G_A(t) = g_A ( 1 + { {2~ t} \over { M_A^2}} ),~ G_P (t) = 4 M_N (
{ { f_{\pi}} g_{\pi N} \over {M_N  }} { 1 \over { m_{\pi}^2 - t }}
- {{2 g_A } \over M_A^2} ),~ G_{\pi N} (t) = g_{\pi N} ( 1 +
\Delta_{\pi N} {  {t - m_{\pi}^2  } \over m_{\pi}^2})~,
\end{equation}
where the GT deviation constant $\Delta_{\pi N} = 0.026$, the
axial mass $M_A$ = 1069 MeV, $g_{\pi N} = 13.21, ~g_A = 1.267,
~f_{\pi} = 92.4 $MeV. It would be also noticeable that it can be
derived from the pre-QCD PCAC hypothesis with a pion source
function \cite{Amal79}.

Of course, under the lowest effective lagrangian, which
corresponds to take $G_P (t)= {{4 M_N^2} \over {m_{\pi}^2 - t }}
g_A $, $G_A (t) = g_A$, and $G_{\pi N} (t) = M_N g_A / f_{\pi}$ by
discarding the terms beyond leading order contributions, the $\Pi
(t)$ can be approximated as ${{m_{\pi}^2 \over { 4 M_N}}} G_P
(t)$. This approximation causes both form factors to have the same
$t$ dependence, and as a result the eq.(\ref{total}) gives the
usual t-channel contribution expressed in terms of the $G_P (t)$.
Moreover the t-channel contribution at the pion threshold turned
out to be composed of two equal contributions from the $G_P (t)$
in $C_a^{\nu}$ and the $\Pi (t)$ in ${{\Sigma}_{a}^{\nu}}$ term,
respectively
\begin{equation}
  {\bar u} (p_2) { I_a \over 2} \epsilon_0 ~[
  { - {2 m_{\pi}} \over f_{\pi} }
  {G_{P}(t) \over { 2M_N}} ]~ {\gamma_5} u(p_{1})
 = {{2 i ~ \epsilon_{\nu}}    \over {
{m_{\pi}} f_{\pi} } } {\Sigma}_{a}^{\nu}  = {{2 \epsilon_\nu
C_a^\nu} \vert_{pole} \over f_{\pi}}~,
\end{equation}
where we used the ${\bf \epsilon} \cdot {\bf \Sigma} $ = 0
\cite{Kama89,Bern91}. It means that the contribution of
$\Sigma_a^0 $ to the transition amplitude for electro-production
occurs as a half of the t-channel contribution to whole $L_0^+$
amplitude at the pion threshold.

Here, we estimate to what extent both form factors affect the
$L_0^+$ amplitude in the pion electro-production near pion
threshold. The transition amplitude for the pion
electro-production is simply given as
\begin{equation}
{ { - e} \over { 4 \pi ( 1 + \mu)} } \epsilon_\mu {\cal M}^{\mu}
\vert_{thr.} = \chi_f^+ [ E_0^+ {{\bf \sigma}} \cdot {\bf b} +
L_0^+ {\bf \sigma} \cdot {\hat {\bf k}} {\hat {\bf k}} \cdot {\bf
a} ] \chi_i ~, \label{sigmaeq28}
\end{equation}
where ${\bf a} = {\bf \epsilon} - { \epsilon_0 \over k_0} {\bf
k}$, ${\bf b} = {\bf a} - {\hat {\bf k}} ( {\hat {\bf k}} \cdot
{\bf a} ) = {\bf \epsilon} - {\hat {\bf k}} ( {\hat {\bf k}} \cdot
{\bf \epsilon} )$ and $\mu = m_{\pi} / M_N$. The result for the
$L_0^+$ amplitude is well known in terms of the Ball amplitudes
$A_i$ which are determined from all pole contributions and the
gauge invariance in the tree diagram approach
\begin{eqnarray}
E_0^+ \vert_{thr.} & = & - { e \over  2M_N} {[ { M_N \over { 4 \pi
W}} {\sqrt {  {E_i  + M_N } \over {2 M_N  }}} ~{\cal A}] }_{thr.}~, \\
\nonumber L_0^+ \vert_{thr.} & = & E_0^+ \vert_{thr.} - { e \over
2M_N} {[ { {E_i - M_N  } \over {2 M_N  }} { M_N \over { 4 \pi W}}
{\sqrt { {E_i  + M_N } \over {2 M_N  }}}~ {\cal B} ] }_{thr.}~,
\end{eqnarray}
where ${\cal A} = A_1 + { \mu \over 2} A_5$ and ${\cal B} = - {1
\over 2 } A_2 + A_4 - A_5 + { 1 \over 4} (2 +  \mu) A_6 - { 1
\over 2} ( 2 + \mu) A_8 $. In the pseudo-threshold limit (${\bf k}
\rightarrow 0$), this result satisfies $E_0^+ = L_0^+$ resulting
from the gauge invariance \cite{Schae91,Amal79}.

Following the tree diagram approach by the effective lagrangian,
our $E_0^+$ and $L_0^+$ amplitudes for $\pi^+$ electro-production
at the pion threshold are given as
\begin{eqnarray}
{E_0^+} (k^2) \vert_{thre.} & = &
  C ~ g_{\pi N} [ { G_A (t) \over g_A} - { \mu \over 2} -
  {  {\nu_2  } \over {2 - \nu2  }}G_M^n ] + O [\mu \nu_2, \mu^2]~,
\\ \nonumber
  {L_0^+} (k^2) \vert_{thre.}& = &
  C ~ g_{\pi N} [ {  m_{\pi}^2 \over {  2 f_{\pi} g_{\pi N}}}
  ({ {G_P (t)  } \over {4 M_N  }} + { {\Pi (t)  } \over {m_{\pi}^2  }}
    ) F_{\pi} (k^2) - { \mu \over 2} -
  {  {\nu_2  } \over {2 - \nu2  }}G_E^n ]+ O [\mu \nu_2, \mu^2]~,
\label{e0+l0+}
\end{eqnarray}
where $ C ={  e  \over { {\sqrt 2} M_N} } { 1 \over { 4 \pi ( 1 +
\mu ) }} \sqrt {  {  { {(2 + \mu)}^2 - ~ \nu_2 } \over { 4 ( 1 +
\mu)  } }}~ $ and $\nu_2 = k^2 / 4 M_N^2$, and the momentum
transfer $t$ is given in terms of $k^2$. The additional ${< r_A
>}^2$ dependent contribution in the contact term is included in $O [\mu \nu_2,
\mu^2]$. As noted in the above t-channel argument, under the GT
relation, our approach reduces to the standard results as those of
ref. \cite{Schae91,Otha89} because the $\Pi (t)$ is approximated
as $ {m_{\pi}^2 G_P (t)  } \over {4 M_N}$ {\it i.e.} ${ {G_P (t) }
\over {4 M_N  }} + { {\Pi (t) } \over {m_{\pi}^2  }} = { {G_P (t)
\over { 2 M_N}}}$.

The phenomenological approach \cite{Amal79,Mainz}, which was
exploited in the extraction of the $L_0^+$ amplitude from
experimental data, can be reproduced if the $L_o^+$ amplitude is
represented in terms of the $\Pi (t)$ {\it i.e.} ${ {G_P (t) }
\over {4 M_N }} + { {\Pi (t) } \over {m_{\pi}^2 }} = { {2 \Pi (t)
\over { m_{\pi}^2}}}$. Our $L_0^+$ amplitude corresponds to the
case of $\lambda = \infty $ in the phenomenological approach
because the $\Pi (t)$ can be decomposed as
\begin{equation}
2 \Pi (t) = { {2 m_{\pi}^2 f_{\pi} g_{\pi N}  } \over {m_{\pi}^2
-t }} + 2 ( g_A M_N - g_{\pi N} f_{\pi} )~.
\end{equation}
The divergence form factor $D (t)$, which was firstly introduced
at ref. \cite{Amal79} by GT relation under the soft pion limit,
equals to 2$\Pi (t)$ in our approach, so that the $D (t)$ is more
natural to be interpreted as the PS form factor. Therefore our
approach reasonably includes both standard and phenomenological
results, and suggests their reciprocal relations under the GT
relation. But, beyond GT relation, both form factors should be
distinguished as in our results. Chiral perturbation calculations
\cite{Ga91,Bern02} have given accurate results for the $E_0^+$ and
$L_0^+$ amplitudes in a systematic way, but the relevant form
factors have been exhausted sometimes by the perturbation itself
in each order.

In order to directly extract the PS form factor from experiments,
one has to separate the t-channel contribution from the whole
amplitude. Such a separation is too hard task to do
experimentally. Even the $L_0^+$ amplitude near threshold has not
yet been reported, despite the try at Saclay \cite{Choi93} and
Mainz \cite{Mainz}.

In theoretical side such a separation is possible. The
contribution of the $\Sigma_a^0 $ term to the $L_0^+$ amplitude is
easily obtained from the above $L_0^+$ amplitude. The ratio $R = {
{L_0^+}_{ (\Sigma_a^0)} / { L_0^+} }$ for $\gamma^* \pi^+$ at the
pseudo-threshold limit (or $\gamma$ point) is about ${ 1 \over
2}$. This means that a half of the total $L_0^+$ amplitude comes
from the $\Sigma_a^0$.

Since this $\Sigma_a^0$ term is closely related to the PS quark
distribution on the nucleon, the $ L_0^+ $ amplitude near
threshold for $\pi^+$ electro-production could give invaluable
information about the nucleon structure, similarly to the role of
the $\sigma$-term in $\pi - N$ scattering on the understanding of
the scalar quark distribution on the nucleon.

Brief summary is done as follows. In pion electro-production, the
t-channel pole in Born terms is explained at the pion threshold as
a sum of two pion poles. The former comes from the induced PS form
factor in the axial current and the latter results from the time
component of the $\Sigma_a^{\nu}$ term, which originates from the
ECSB effect, likewise the $\sigma_{\pi N}$ term in $\pi - N$
scattering. In principle, the $t$ dependence of both form factors
is fully different, so that the unique extraction of the $G_P(t)$
from the experimental data leaves another ambiguity due to the PS
form factor $\Pi (t)$. More thorough investigation is necessary
because not only the $G_P (t)$ and $F_{\pi} (k^2)$ but also the
$\Pi (t)$ are involved in the extraction of $L_0^+$ amplitude from
the experimental data.

But, under the lowest effective lagrangian which maintains the GT
relation, both form factors show the same $t$ dependence, and as a
result the sum was shown to equally account for the t-channel in
the tree diagram approach. Our analysis shows that a half of the
$L_0^+$ amplitude in $\pi^+$ electro-production is attributed to
the contribution due to the $\Sigma_a^0$ term, which represents a
charged PS quark density on the nucleon. We estimated its value as
38 MeV. In this paper, only the charged PS quark-density on the
nucleon is discussed. Weak pion production could be a good guide
to the neutral PS quark-density distribution on the nucleon.

This work was supported by the Soongsil University Research Fund.

\def\prl#1{Phys.\ Rev.\ Lett.\ {\bf #1}}
\def\pr#1{Phys.\ Rev.\ {\bf #1}}
\def\prc#1{Phys.\ Rev.\ C {\bf #1}}
\def\prd#1{Phys.\ Rev.\ D {\bf #1}}
\def\np#1{Nucl.\ Phys.\ {\bf #1}}
\def\pl#1{Phys.\ Lett.\ {\bf #1}}
\def\prt#1{Phys.\ Reports.\ {\bf #1}}
\def\jpsj#1{J.\ Phys.\ Soc.\ Jpn.\ {\bf #1}}

\end{document}